\documentclass[aps,twocolumn,superscriptaddress,showkeys,showpacs]{revtex4-1}
\usepackage{amssymb}
\usepackage{amsfonts}
\usepackage{amsmath}
\usepackage{amsthm}
\usepackage{epsfig}
\usepackage{color} 
\usepackage{subfigure}

\newcommand{\eps}{\varepsilon}

\newcommand{\fr}{\frac}

\graphicspath{{./}{./Figures/}}

\begin{document}

\title{Robustness of chimera states for coupled FitzHugh-Nagumo oscillators}

\author{Iryna Omelchenko}
\email[corresponding author: ]{omelchenko@itp.tu-berlin.de}
\affiliation{Institut f{\"u}r Theoretische Physik, Technische Universit{\"a}t Berlin, Hardenbergstra\ss{}e 36, 10623 Berlin, Germany}
\author{Astero Provata}
\author{Johanne Hizanidis}
\affiliation{Institute of Nanoscience and Nanotechnology, National Center for Scientific Research ``Demokritos'', 15310 Athens, Greece}
\author{Eckehard Sch{\"o}ll}
\affiliation{Institut f{\"u}r Theoretische Physik, Technische Universit{\"a}t Berlin, Hardenbergstra\ss{}e 36, 10623 Berlin, Germany}
\author{Philipp H{\"o}vel} 
\affiliation{Institut f{\"u}r Theoretische Physik, Technische Universit{\"a}t Berlin, Hardenbergstra\ss{}e 36, 10623 Berlin, Germany}
\affiliation{Bernstein Center for Computational Neuroscience, Philippstra{\ss}e 13, 10115 Berlin, Germany}

%\date{\today}

\begin{abstract}
Chimera states are complex spatio-temporal patterns that consist of coexisting domains of spatially coherent and incoherent dynamics. This counterintuitive phenomenon was first observed in systems of identical oscillators with symmetric coupling topology. Can one overcome these limitations? To address this question, we discuss the robustness of chimera states in networks of FitzHugh-Nagumo oscillators. Considering networks of inhomogeneous elements with regular coupling topology, and networks of identical elements with irregular coupling topologies, we demonstrate that chimera states are robust with respect to these perturbations, and analyze their properties as the inhomogeneities increase. We find that modifications of coupling topologies 
cause qualitative changes of chimera states: additional random links induce a shift of the stability regions in the system parameter plane,  gaps in the connectivity matrix result in a change of the multiplicity of incoherent regions of the chimera state, and hierarchical geometry in the connectivity matrix induces nested coherent and incoherent regions.
\end{abstract}

% Synchronization, nonlinear dynamics, 05.45.Xt
% Complex systems, 89.75.-k

\maketitle

\section{Introduction}
Synchronization in systems of coupled oscillators has been 
investigated in various fields such as nonlinear dynamics, network science, and
statistical physics, and found its applications in physics, biology, and technology. 
Fascinating dynamical scenarios, which emerge for coupled systems, have attracted much attention in the 
scientific community \cite{PIK01,BOC06a}.
For instance, a very peculiar type of dynamics called chimera states, 
when a network exhibits a hybrid state combining both coherent and
incoherent parts, was first reported for coupled phase oscillators~\cite{KUR02a,ABR04}.
The surprising aspect of this phenomenon is that these
states were detected in systems of identical oscillators coupled in a
symmetric ring topology with a symmetric interaction
function, and coexist with a completely synchronized state. The last decade has seen an increasing interest in chimera states~\cite{LAI09,MOT10,MAR10,OME10a,OME12a,MAR10b,WOL11a,BOU14}. It was shown that they are not limited to phase oscillators, but can be found
in a large variety of different systems including time-discrete maps~\cite{OME11}
and time-continuous chaotic models~\cite{OME12}, and neural systems~\cite{OME13,HIZ13}. 
Moreover, chimera states were found in systems with higher spatial dimensions~\cite{OME12a,SHI04,PAN13,PAN14}.
Together with initially reported chimera states, which consist of one coherent and one incoherent domain, new types of these peculiar states having multiple incoherent regions~\cite{SET08,OME13}, as well as amplitude-mediated~\cite{SET13,SET14}, and pure amplitude chimera and chimera death states~\cite{ZAK14} were discovered.

Potential applications of chimera states in nature include the phenomenon of unihemispheric sleep in birds and dolphins~\cite{RAT00}, bump states in neural systems~\cite{LAI01,SAK06a}, in power grids~\cite{FIL08}, or in social systems~\cite{GON14}.  
Many works considering chimera states were mostly based on numerical results. A deeper bifurcation analysis~\cite{OME13a} and even a possibility to control chimera states~\cite{SIE14c} were obtained only recently.

Ten years after the first numerical observation, the experimental verification of chimera states was demonstrated in    
chemical~\cite{TIN12,NKO13} and  optical~\cite{HAG12} systems. Further experiments involved  mechanical~\cite{MAR13}, electronic~\cite{LAR13}, and electrochemical~\cite{SCH14a, WIC13} oscillator systems.

Identical elements and symmetric coupling topology were widely assumed to be the necessary ingredients for chimera states. Recent studies show that one can overcome these limitations and chimera-like states can be found also when the elements of the system are nonidentical~\cite{LAI10}, or when the topology is not regular~\cite{KO08,SHA10,LAI12,YAO13,ZHU14} or even global \cite{SCH14a,YEL14}. Then, the spatial order in the system is lost and the assignment to coherent or incoherent regions is based, for instance, on the dynamical variables. Hence, the intriguing phenomena of chimera states continues to show new aspects and is far from being completely explored.    

In the present study, we discuss the robustness of chimera states in systems of nonlocally coupled FitzHugh-Nagumo (FHN) oscillators \cite{FIT61,NAG62} studied in Ref.~\cite{OME13}. We consider two types of inhomogeneities. The first type is realized by inhomogeneous parameters of the local elements in a regular ring topology. As a second type, we study systems with identical units, but irregular coupling topologies. We find that for small inhomogeneities, chimera states are robust under these perturbations, and we discuss qualitative changes of the chimera states caused by different types of irregularities.

The motivation for studying irregular coupling topologies comes from recent results in the area of neuroscience. 
Diffusion Tensor Magnetic Resonance Imaging (DT-MRI) studies revealed an intricate
architecture in the neuron interconnectivity of the human and mammalian brain, which has
already been used in simulations ~\cite{VUK14}. The analysis of DT-MRI images (resolution of the order of $0.5$ mm)
has shown that the connectivity of the neuron axons network represents a hierarchical
geometry with fractal dimensions varying between 2.3 and 2.8, depending 
on the local properties, on the subject, and on the noise reduction
threshold \cite{KAT09,EXP11,PRO12}. Based on these findings, we study the development
of chimera states in coupled neurons operating in
networks which involve connectivity gaps 
as well as in topologies with hierarchical connectivity.
In both cases, we assess the influence of
the irregular connectivity on the properties of the
chimera state.

The rest of the paper is organized as follows: In the next section we introduce the model, a set of $N$ nonlocally
coupled FHN oscillators, and we briefly review the 
conditions for the appearance of chimera states. In Sec.~\ref{sec:inhomo},
we introduce an inhomogeneity in the system parameters via randomly chosen frequencies of the local units, and discuss the robustness of chimera and multi-chimera states. In Sec.~\ref{sec:randomlinks}, we include additional random links on top of the regular nonlocal coupling scheme and obtain stability regions for chimera states depending on the probability of the introduction of new links.  
In Sec.~\ref{sec:gaps}, we consider symmetric and asymmetric gaps in the nonlocal coupling between
each node and its neighbors. Numerical results show that when gaps are introduced, the multiplicity of 
the chimera states may change depending on the position of the gaps. 
Hierarchical geometry in the connectivity between each node and all other nodes
of the network is discussed in  Sec.~\ref{sec:fractal}. Multi-chimeras with nested shapes appear as a result of hierarchical connectivity and appropriate choice of parameters. The main 
conclusions are summarized in Sec.~\ref{sec:concl} and open problems are
discussed.

\section{The model}
\label{sec:model}
We consider a 1-dimensional ring of $N$ nonlocally coupled FHN oscillators, where each element is coupled to $R$ neighbors on either side \cite{OME13}:
\begin{subequations}\label{System:FHN}
\begin{align}
\varepsilon \frac{d u_k}{dt} & = u_k - \fr{u_k^3}{3} - v_k \nonumber\\
                     & + \frac{\sigma}{2R} \sum\limits_{j=k-R}^{k+R}
\left[ b_{\mathrm{uu}}( u_j - u_k ) + b_{\mathrm{uv}}( v_j - v_k ) \right],\\
\frac{d v_k}{dt} &= u_k + a_k \nonumber\\
                     & + \frac{\sigma}{2R} \sum\limits_{j=k-R}^{k+R}
\left[ b_{\mathrm{vu}}( u_j - u_k ) + b_{\mathrm{vv}}( v_j - v_k ) \right],
\end{align}
\end{subequations}
where $u_k$ and $v_k$ are the activator and inhibitor variables, respectively \cite{FIT61,NAG62}, and $\varepsilon>0$ is a small parameter characterizing a timescale separation, which we fix at $\varepsilon = 0.05$ throughout the paper. 
$\sigma$ denotes the coupling strength. It is convenient to consider the ratio $r=R/N$, called coupling radius, which ranges from $1/N$ (nearest-neighbor coupling) to 0.5 (all-to-all coupling). All indices are modulo $N$. 
Depending upon the threshold parameter $a_k$, $k=1,\cdots,N$, each individual FHN unit exhibits either oscillatory ($|a_k|<1$) or excitable ($|a_k|>1$) behavior. In this study, we assume that the elements are in the oscillatory regime, i.e., $a_k\in(-1,1)$.

The important feature of Eqs.~(\ref{System:FHN}a,b) is that they contain not only direct $u$-$u$ and $v$-$v$ couplings,
but also cross-couplings between activator ($u$) and inhibitor ($v$) variables, which we model by a rotational coupling matrix:
\begin{eqnarray}
\mathbf{B} = \left(
\begin{array}{ccc}
b_{\mathrm{uu}} & & b_{\mathrm{uv}} \\
b_{\mathrm{vu}} & & b_{\mathrm{vv}}
\end{array}
\right) =
\left(
\begin{array}{ccc}
\cos \phi  & & \sin \phi \\
-\sin \phi  & & \cos \phi
\end{array}
\right).
\label{Matrx:B}
\end{eqnarray}
Therefore, the matrix $\mathbf{B}$ is determined by the coupling phase $\phi$.

The existence of chimera states for the system~(\ref{System:FHN}a,b) with identical elements $a_k\equiv a, k=1,\cdots,N$ was reported in Ref.~\cite{OME13}. There, applying a phase-reduction technique~\cite{IZH00b}, we identified nonlocal off-diagonal coupling to be a crucial ingredient for the occurrence of chimera states in the system. In addition, multi-chimera states consisting of multiple domains of incoherence were observed.

%*******************************************************
\begin{figure*}[ht!]
\includegraphics[width=\linewidth]{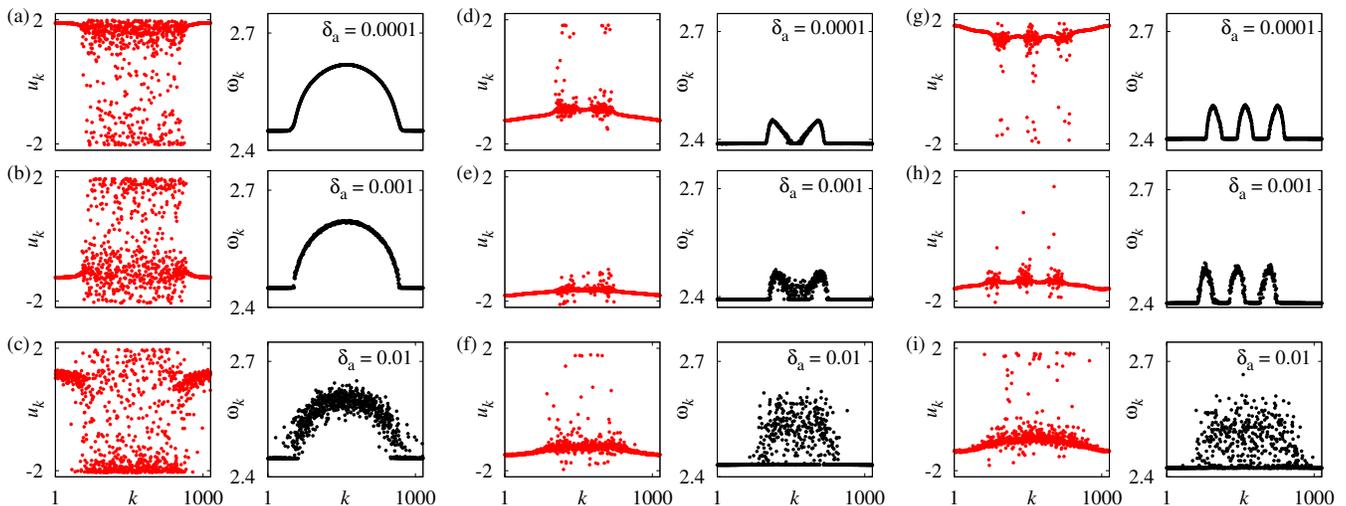}
\caption{(Color online) Snapshots of the variables $u_k$ and mean phase velocities $\omega_k$ for inhomogeneous oscillators. (a),(b),(c) $r=0.35,$ $\sigma=0.1$; (d),(e),(f) $r=0.33,$ $\sigma=0.28$;  (g),(h),(i) $r=0.25,$ $\sigma=0.25$; values of $\delta_a$ shown for each panel above the mean phase velocities plots. System parameters: $N=1000$, $\phi=\pi/2-0.1$, $a_{\text{mean}}=0.5$, and $\eps=0.05$.}
\label{fig1}
\end{figure*}
%*******************************************************

\section{Inhomogeneous elements}
\label{sec:inhomo}
In this section we consider system ~(\ref{System:FHN}a,b) in the case when the local FHN units are nonidentical and have different frequencies. This feature can be achieved by random choice of the parameters $a_k$, because the parameter $a_k$ determines the period of the individual element \cite{BRA09}. Our attention is focused on the case when the threshold parameters $a_k$ are drawn from a normal (Gaussian) distribution with mean value $a_{\text{mean}}$ and variance $\delta_a$.  In our numerical simulations, we have also considered the case when  the threshold parameters $a_k$ were drawn from a uniform random distribution, and obtained qualitatively similar results.

Together with the coupling radius $r$, coupling strength $\sigma$, and phase $\phi$ of the rotational local interaction matrix, the parameters  $a_{\text{mean}}$ and  $\delta_a$ now control the system dynamics.
In the following, we choose $a_{\text{mean}} =0.5$, and vary the variance $\delta_a$. This quantity will define the amount of inhomogeneity of the elements in our system. To compare our results with the system of identical elements, we will follow the choice of  system parameters as considered in Ref.~\cite{OME13}.

A significant feature of chimera states is the difference of average frequencies of the oscillators that belong to the coherent or incoherent parts, respectively. This feature can be visualized by the mean phase velocities for each oscillator calculated as $\omega_k = 2 \pi M_k/\Delta T$, $k=1,\cdots,N$, where $M_k$ denotes the number of oscillations and $\Delta T$ is the simulation time.
The profile of the mean phase velocities is usually characterized by a plateau of equal frequencies that correspond to the coherent domain of the chimera state, and an arc-like part corresponding to the incoherent part. 

Figure~\ref{fig1} shows typical snapshots and corresponding profiles of the mean phase velocity for chimera states with one, two, and three incoherent regions, as the inhomogeneity in the system increases. The number of incoherent regions depends on the coupling strength $\sigma$ and coupling range $r$ \cite{OME13}. Starting the simulation for each panel with initial conditions that are randomly distributed on the circle $u_k(0)^2+v_k(0)^2=4$, we obtain chimera states with well pronounced mean phase velocity profiles that demonstrate a clear difference between coherent and incoherent domains. When the system elements have very close frequencies and the variance is small $\delta_a=0.0001$ [Fig.~\ref{fig1}(a),(d),(g)], chimera states with one and multiple incoherent regions are robust. The difference $\Delta \omega$ between the maximum mean phase velocity of the incoherent domain
and the $\omega$-value of the coherent domain is larger in the single chimera state than in the multi-chimera
states. 
For a small increase of the inhomogeneity, e.g. $\delta_a=0.001$ [Fig.~\ref{fig1}(b),(e),(h)], the chimera state with one incoherent domain remains visible, but the profiles of the mean phase velocity for chimera states with two and three incoherent domains become more noisy around the incoherent parts.  For even larger inhomogeneities such as $\delta_a=0.01$ [Fig.~\ref{fig1}(c),(f),(i)], the multiple incoherent domains merge into one larger region. However, the corresponding mean phase velocity profile does not resemble that of the original single
chimera state [Fig.~\ref{fig1}(c)], which maintains the arc-shaped form of the homogeneous system. For the chimera state with one incoherent domain we still clearly observe the difference of the mean frequencies for incoherent and coherent domain. These observations can be explained as follows: In multi-chimera states frequencies are less distinguishable (smaller $\Delta \omega$). Therefore, introducing inhomogeneities $\delta_a$ leads to the collapse of the $\omega$ profile to a noisy configuration, where incoherent oscillators take over the coherent 
ones.

In the numerical simulations of the system~(\ref{System:FHN}a,b) for fixed values of $a_{\text{mean}}$ and $\delta_a$, we use different realizations of the threshold parameters $a_k, k=1,\cdots,N.$ Moreover, as we consider randomly chosen initial conditions, it is important to understand how their choice influences the obtained system solutions. 

%*********************************************************
\begin{figure*}[ht!]
\includegraphics[height=0.75\linewidth, angle =270]{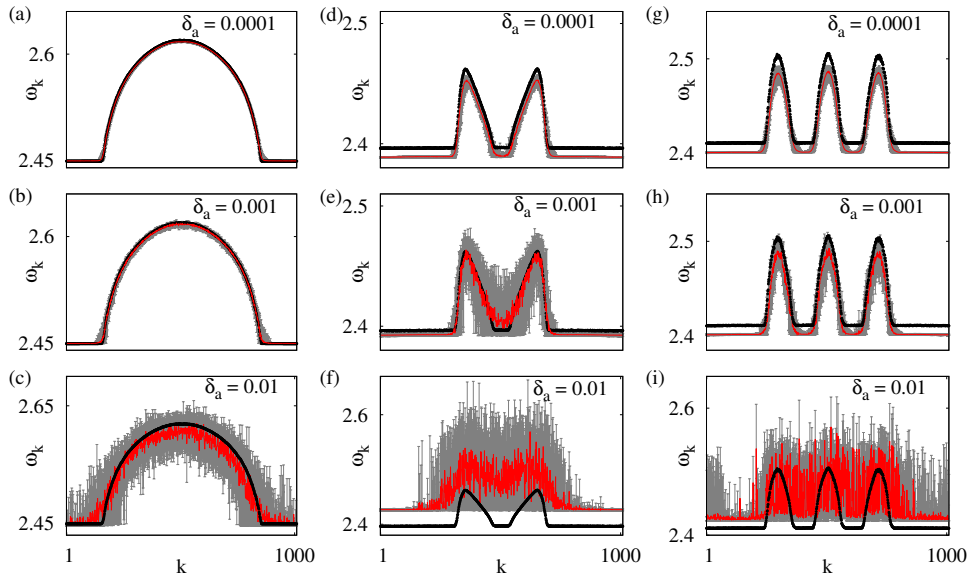}
\caption{(Color online) Averaged mean phase velocities for chimera states with one and multiple incoherent domains with increasing system inhomogeneity. Black denotes phase velocities for the system of identical oscillators $a_k=0.5, k=1,\cdots,N$, red (gray)- mean phase velocities averaged over $100$ realizations, gray (light gray)- variations of phase velocities $(\omega_{\text{min}}, \omega_{\text{max}})$ in $100$ realizations. (a),(b),(c) $r=0.35,$ $\sigma=0.1$; (d),(e),(f) $r=0.33,$ $\sigma=0.28$;  (g),(h),(i) $r=0.25,$ $\sigma=0.25$; values of $\delta_a$ are shown inside each panel. Other system parameters as in Fig.~\ref{fig1}.}
% : $N=1000$, $\phi=\pi/2-0.1$, $a_{\text{mean}}=0.5$, and $\eps=0.05$.}
\label{fig2}
\end{figure*}
%*********************************************************

To address this issue, Fig.~\ref{fig2} depicts the comparison of mean phase velocities for chimera states with one, two, and three incoherent domains with increasing of the system inhomogeneity. 
Red (gray) lines in Fig.~\ref{fig2} show the averaged mean phase velocities over $100$ realizations, and gray (light gray) error bars denote the variations of the obtained values, which is given by the minimum and maximum. Black lines show the values of mean phase velocities for system with identical elements $a_k=a_{\text{mean}}=0.5$ for comparison.

For a small value of $\delta_a=0.0001$ [Fig.~\ref{fig2}(a),(d),(g)], we obtain mean phase velocity profiles with pronounced differences between the coherent and incoherent domains and the variations of the velocities are in general small.
As inhomogeneity increases, $\delta_a=0.001$ [Fig.~\ref{fig2}(b),(e),(h)],
the variations for mean phase velocities of multi-chimera states become more pronounced, although the average value still shows maxima for incoherent domains. For larger inhomogeneities, $\delta_a=0.01$ [Fig.~\ref{fig2}(c),(f),(i)], the variations are strong and appear even in the coherent domain and multiple incoherent regions collapse into one. The chimera state with one incoherent domain, however, appears to be more robust than in the case of multi-chimera.

The oscillation period of each FHN system strongly depends on the value of its threshold parameter $a_k$~\cite{BRA09}. With a larger range for the values of $a_k$, we increase the spectrum of their frequencies, which are then additionally influenced by the coupling term. This results in the changes of the  individual frequencies of the oscillators in the inhomogeneous system. 
Therefore, the mean phase velocities for the oscillators that belong to the coherent domains of multi-chimera states are smaller and then become larger compared to the case of identical elements when the value of $\delta_a$ increases. See black and gray (red) curves in Fig.~\ref{fig2}.
 
To summarize this section, inhomogeneous system parameters in rings of nonlocally coupled FHN oscillators still allow us to observe chimera states. Classical chimera states with one incoherent domain are more robust under this perturbation than multi-chimera states with multiple incoherent domains. Chimera states can be observed only for small inhomogeneities. Otherwise, their incoherent parts merge and they form a chimera state with a single, large incoherent part.

\section{Irregular topology: additional random links}
\label{sec:randomlinks}
Studies of the connectivity in neural networks show that local couplings often coexist with long-range connections between the neurons \cite{SCA95,HEL00,HOL03}. In such coupling structures, the interplay between the local interactions and long-range shortcuts can give birth to nontrivial dynamical effects.
Starting from this idea, we consider in this section a perturbation of the regular ring coupling of system~(\ref{System:FHN}a,b). Keeping the original coupling radius, we add new links, i.e. shortcuts, with  probability $p$ between the elements. No multiple edges between the same nodes are allowed. The transformed system can be written as

\begin{subequations}\label{System:RandLinks}
\begin{align}
\varepsilon \frac{d u_k}{dt} & = u_k - \fr{u_k^3}{3} - v_k \nonumber\\
                     & + \frac{\sigma}{2R+L_k} \sum\limits_{j=1}^{N} C_{kj}\left[ b_{\mathrm{uu}}( u_j - u_k ) + b_{\mathrm{uv}}( v_j - v_k ) \right],\\
\frac{d v_k}{dt} &= u_k + a_k \nonumber\\
                     & + \frac{\sigma}{2R+L_k} \sum\limits_{j=1}^{N} C_{kj}\left[ b_{\mathrm{vu}}( u_j - u_k ) + b_{\mathrm{vv}}( v_j - v_k ) \right],
\end{align}
\end{subequations}
where again 
% $u_k$ and $v_k$ are the activator and inhibitor variables, and $\varepsilon>0$ is a small parameter fixed at $\varepsilon = 0.05$. 
$\sigma$ denotes strength of the coupling. All indices are modulo $N$ and the individual FHN elements are in the oscillatory regime and identical, i.e., $a_k\equiv a = 0.5$.
% In the next two sections  we fix $a=0.5$. 
$\mathbf{C}=\left\{C_{kj}\right\}_{k,j=1,\cdots,N}$ is the adjacency matrix with elements $0$ or~$1$, $R$ denotes the number of nearest neighbors, and $L_k$ is the number of new links added to the $k$-th element, which is $L_{k} \approx (N-2R)p$. 
The number of connections for each element increases for larger probability $p$. Thus, we rescale the coupling strengths by the updated number of neighbors such that every connection has the same strength, for
comparison with the regular ring topology of system~(\ref{System:FHN}a,b) \cite{OME13}. 

The introduction of shortcuts shares similarities to a well-known small-world network~\cite{WAT98} with one difference. Here, we do not rewire already existing links when adding new ones, and the regular nonlocal structure is kept. For large networks, both strategies lead to similar topologies \cite{NEW99b}.

The dynamics of the system~(\ref{System:RandLinks}a,b) is defined by five control parameters: the threshold parameter $a$ of the individual uncoupled FHN unit, the strength of the coupling $\sigma$, the phase $\phi$ of the rotational matrix $\mathbf{B}$ defining local interaction scheme, the regular coupling radius $r$, and the probability $p$ of adding new links. The last two parameters give us an approximate number of interactions for each system element, that is, an {\it effective} coupling radius $\tilde{r}$, which can be calculated as 
\begin{equation}\label{CouplRadiusEffective}
\tilde{r} = \dfrac{R + (N/2-R)p}{N}.  
\end{equation}

%*********************************************************
\begin{figure}[t!]
\includegraphics[height=\linewidth, angle=270]{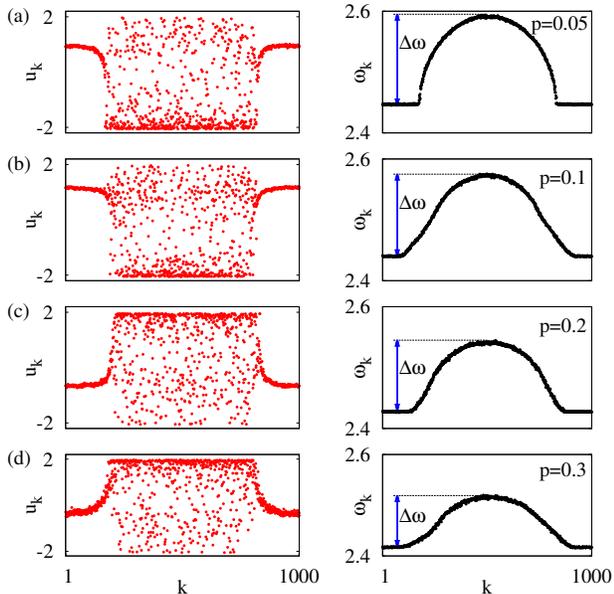}
\caption{(Color online) Snapshots of the variables $u_k$ (left panels) and mean phase velocities (right panels) for the system Eqs.~(\ref{System:RandLinks}a,b). Probability of adding new links: (a)~$p=0.05$, (b)~$p=0.1$, (c)~$p=0.2$, (d)~$p=0.3$. Parameters: coupling radius $r=0.35$, coupling strength $\sigma=0.1$, $a_k=a=0.5$, and other parameters as in Fig.~\ref{fig1}.}
% $N=1000$, $r=0.35,$ $\sigma=0.1$, $\phi=\pi/2-0.1$, $a=0.5$, and $\eps=0.05$.}
\label{fig3}
\end{figure}
%*********************************************************

Figure~\ref{fig3} shows examples of chimera states obtained in the system~(\ref{System:RandLinks}a,b) for intermediate regular coupling radius $r=0.35$ and small coupling strength $\sigma=0.1$, with increasing probability for new links. This parameter choice corresponds to chimera states in the regularly coupled system with no additional links shown in Fig.~\ref{fig1}(a). Left and right panels in Fig.~\ref{fig3} depict snapshots and mean phase velocity profiles, respectively. Looking at the snapshots, chimera states appear to be robust in system~(\ref{System:RandLinks}a,b). Increasing the number of new links results in the following changes in the phase velocity profiles: The characteristic arc-like shape of the incoherent region is still visible, but the border to the coherent domain becomes less sharp. Furthermore, we find that the difference between the maximum and minimum of the average phase velocities,  $\Delta\omega = \omega_{\text{max}} - \omega_{\text{min}}$, decreases.
 This effect can be explained by considering the 
additional input by the oscillators of the coherent part applied to the oscillators of the incoherent part. Elements, which had a larger difference in the phase velocities in the unperturbed case (Sec.~\ref{sec:inhomo}), are now in contact. This leads to homogenization of their frequencies and thus to a smaller $\Delta\omega$.  

Chimera states with multiple incoherent regions are much more sensitive with respect to this topological perturbation of the regular ring topology. Even a small number of shortcuts results in the loss of smaller coherent domains in between the incoherent ones (cf. Figs.~\ref{fig1} and \ref{fig2}). This leads to the formation of a chimera state with one larger incoherent part (results not shown).

%*********************************************************
\begin{figure*}[ht!]
\includegraphics[height=\linewidth, angle=270]{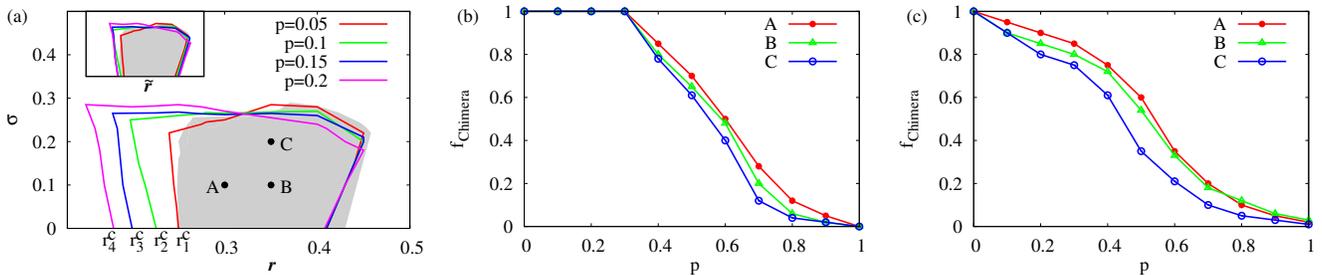}
\caption{(Color online) (a) Stability regions for chimera states with one incoherent domain in the plane of coupling radius~$r$ and coupling strength~$\sigma$. Gray region corresponds to the system with regular nonlocal coupling topology ($p=0$), and lines denote the borders of stability region for $p=0.05, 0.1, 0.2, 0.3$. The inset shows the same plots rescaled to the effective coupling radius~$\tilde{r}$. (b) Fraction $f_{\text{Chimera}}$ of successful realizations of chimera state starting from random initial conditions in $500$ numerical experiments for each parameter set. The parameter values denoted by black dots in panel~(a) are: $A~(r=0.3, \sigma=0.1)$, $B~(r=0.35, \sigma=0.1)$, and $C~(r=0.35, \sigma=0.2)$. Other parameters as in Fig.~\ref{fig1} with $a_k=a=0.5$.
% $N=1000$, $\phi=\pi/2-0.1$, $a=0.5$, and $\eps=0.05$.
(c)~Fraction $f_{\text{Chimera}}$ of successful realizations of chimera state in the network with rewired links, all parameters as in panel~(b).}
\label{fig4}
\end{figure*}
%*********************************************************

Figure~\ref{fig4}(a) depicts the stability regions for chimera states with increasing probability for random shortcuts. The gray region shows the stability region of chimera state with one incoherent domain in the original system~(\ref{System:FHN}a,b) without additional random links, reproduced from Ref.~\cite{OME13} as a reference case. The increase in the number of random links in the network results in a transformation of the stability region: the regions change their form and their left border moves in the direction of smaller coupling radii. Hence, additional random shortcuts allow for the existence of chimera states for smaller regular coupling radius, where they were not possible in the system with regular coupling topology. 

The left border of the stability regions for different values of $p$, which can be conceived as the critical coupling radius, is given by $r_i^c$ [see Fig.~\ref{fig4}(a)]. According to Eq.~(\ref{CouplRadiusEffective}), the corresponding values of the critical effective coupling radius are
$\tilde{r}_{p=0.05}^{\text{crit}} (r_1^c) \approx 0.262$,
$\tilde{r}_{p=0.1}^{\text{crit}} (r_2^c) \approx 0.253$,
$\tilde{r}_{p=0.15}^{\text{crit}} (r_3^c) \approx 0.245$, and 
$\tilde{r}_{p=0.2}^{\text{crit}} (r_4^c) \approx 0.244$. 
They appear to be close to the critical coupling radius of the regular system: $\tilde{r}_{p=0}^{\text{crit}} = 0.25$. In addition, we observe a data collapse by rescaling to the effective coupling radius [see inset of Fig.~\ref{fig4}(a)]. 

Due to the presence of the randomness in the system, we can statistically analyze the persistence of chimeras. More precisely, we investigate the probability to obtain a chimera state starting from random initial conditions with random realization of the system topology for a fixed value of probability $p$. Let us fix three points inside the stability region for the unperturbed regular system $A~(r=0.3, \sigma=0.1)$, $B~(r=0.35, \sigma=0.1)$, and $C~(r=0.35, \sigma=0.2)$. Figure~\ref{fig4}(b) shows the fraction $f_{\text{Chimera}}$ of successful realizations of chimera states, which we obtained starting from random initial conditions and random setup of shortcuts over 500 realizations for each parameter set.
For a small number of shortcuts, we always obtain a chimera state. This also confirms the fact that points A, B, and C remain inside the stability regions shown in Fig.~\ref{fig4}(a). With increasing $p$ and correspondingly a larger number of shortcuts, the fraction of realizations with chimera states decreases. The behavior of the system becomes highly multistable and the basin of attraction for chimera states shrinks. Surprisingly, even for very large number of shortcuts, there is still a small probability to obtain a chimera state in the system,
although the coupling is close to all-to-all.

For comparison, let us consider now a classical small-world network~\cite{WAT98}, where the existing links are rewired with probability $p$ to a random node. In this case, the effective coupling radius will coincide with the coupling radius of the original regular network. The probability $p$ of adding random links serves again as a measure of the network irregularity. Figure~\ref{fig4}(c) depicts the fraction $f_{\text{Chimera}}$ of successful realizations of chimera states over 500 numerical realizations in a similar way as in Fig.~\ref{fig4}(b), and for the same system parameters.  Even for a small amount of random links, only a part of the realizations lead to chimera states, in contrast to the previous case. This can be explained by the fact that rewiring of the existing links  destroys the 
strong effect of the nonlocal coupling kernel, still present in networks in which the regular ring is maintained as a backbone. Increasing the probability $p$, the  irregularity in the coupling  topology becomes stronger, and the basin of attraction for chimera states decreases. 

We conclude that adding of random links to the system of nonlocally coupled FHN oscillators still allows us to observe chimera states. For small probabilities, we are able to obtain stability regions for chimera states in the case when we do not rewire already existing links, and the effect of nonlocal coupling kernel in the system predominates. For larger numbers of additional links, the basin of attraction of chimera states decreases rapidly, although chimera states still can be observed even for a large amount of added random links, in both cases when the existing links are rewired or kept.

The numerical simulations presented in Fig.~\ref{fig4}(b),(c) were obtained for systems of $N=1000$ elements. We performed analogous simulations for other systems sizes, and obtained similar results for larger systems ($N=5000, 10000$). For smaller system sizes ($N=100, 200$) we found smaller fractions of successful realizations of chimera states. This can be explained by the fact that for smaller systems chimera states perform a random spatial motion of their coherent/incoherent regions (drift)~\cite{OME10a}. Then, the averaged mean phase velocity profiles do not show the characteristic arc shape, and chimera states are barely detectable.

%**************************************************************************
%**************************************************************************
%**************************************************************************

\section{Homogeneous and Inhomogeneous Gap Distributions}
\label{sec:gaps}
After the investigation of shortcuts added to the regular ring, we will address another modification of the topology~Eqs.~(\ref{System:RandLinks}a,b) in this section that gives rise to a different form of inhomogeneities. 
These are associated with the presence of gaps in the connectivity matrix. This is inspired 
by current research in neuroscience, and in particular from  functional magnetic resonance imaging (fMRI) task 
experiments used to
identify connectivity between different parts of the brain. Early fMRI investigations
have shown that for simple tasks, such as parroting or finger tapping,
only some parts of the brain show coherent electromagnetic activity, while the 
rest shows normal disordered activity. This observation indicates that the overall
connectivity is fragmented with gaps in the connectivity matrix. In other words,
each node has a preferable attachment to some groups of nodes and a weak
attachment to others.

Before introducing inhomogeneities in the connectivity matrix, we consider the question 
if synchronization is related to the symmetry in the link arrangement
around each node. To answer this question we create an asymmetric, directed
network, where each node is linked
only to its right neighbors. We compare this symmetry-broken configuration with the case of a symmetric connectivity
network, keeping all other parameters fixed.
For notational convenience, the node $k$, around 
which the connectivity is described, will be termed {\it reference node}.
Next, we consider the case where the reference node is not linked to its direct $2R$ neighbors,
but to neighbors $G$ positions away. For one-sided connectivity, node $k$ is linked
to nodes $k+G+1, k+G+2, \cdots, k+G+2R$. 
For comparison with previous studies \cite{OME13}, we consider the following set of
parameters: $N=1000$, $\varepsilon =0.05$, $a =0.5$, $\sigma =0.28$, $\phi=\pi/2-0.1$, and $R=330$, which has been studied earlier \cite{OME13}. 

%***********************************************************************************
\begin{figure}[ht!]
%\begin{centering}
\centering
\includegraphics[clip,width=0.38\textwidth,angle=-90]{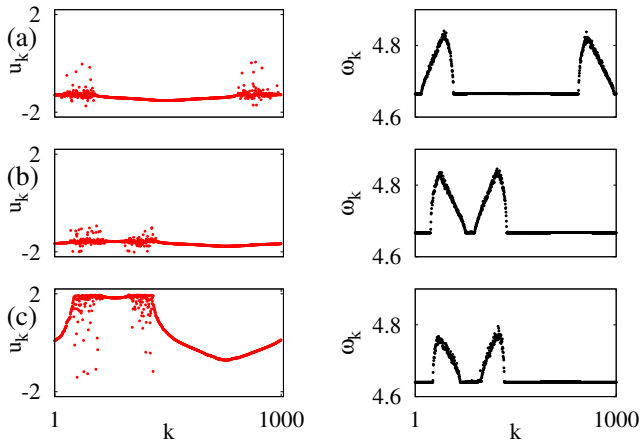}
%\end{centering}
\caption{\label{fig:5} (Color online) Snapshots 
and corresponding mean phase velocity profiles for different connectivity distributions.
(a) Symmetric connectivity distribution $R_L=330$, $R_R=330$,
(b) asymmetric connectivity distribution $R_L=0$, $R_R=660$,
and (c) displaced asymmetric connectivity distribution, $R_L=0$, $G=200$,
$R_R=660$. Other parameters are: 
$N=1000$, $\varepsilon =0.05$, $a =0.5$, $\phi=\pi/2-0.1$, $\sigma =0.28$. All simulations start from the same initial conditions.}
\end{figure}
%**********************************************************************************

Figure~\ref{fig:5} shows the snapshots and mean phase velocity profiles
for the following three cases: (a) the links are symmetrically placed around each node $R_L=R=330$ to the left and 
 $R_R=R=330$ to the right, (b) all $R_R=2R=660$ links are placed to the right of each node
and $R_L=0$, (c) all $R_R=2R=660$ links are placed to the right of each node [as in case (b)] but there is a displacement
$G=200$ in the connectivity of each node. Note that the overall number of neighbors remains constant at $2R=660$ for each node.

From Fig.~\ref{fig:5} it is obvious that the multiplicity
of the chimera state does not change qualitatively.
This result suggests that symmetry around the reference node is not essential for the formation of chimeras
and has been verified for other coupling radii (results not shown).
These observations do not depend on
the specific arrangement of the links as long as they are densely packed (without gaps).
Effects of gaps are discussed next.

\begin{figure}[ht!]
\centering
\includegraphics[width=0.85\linewidth]{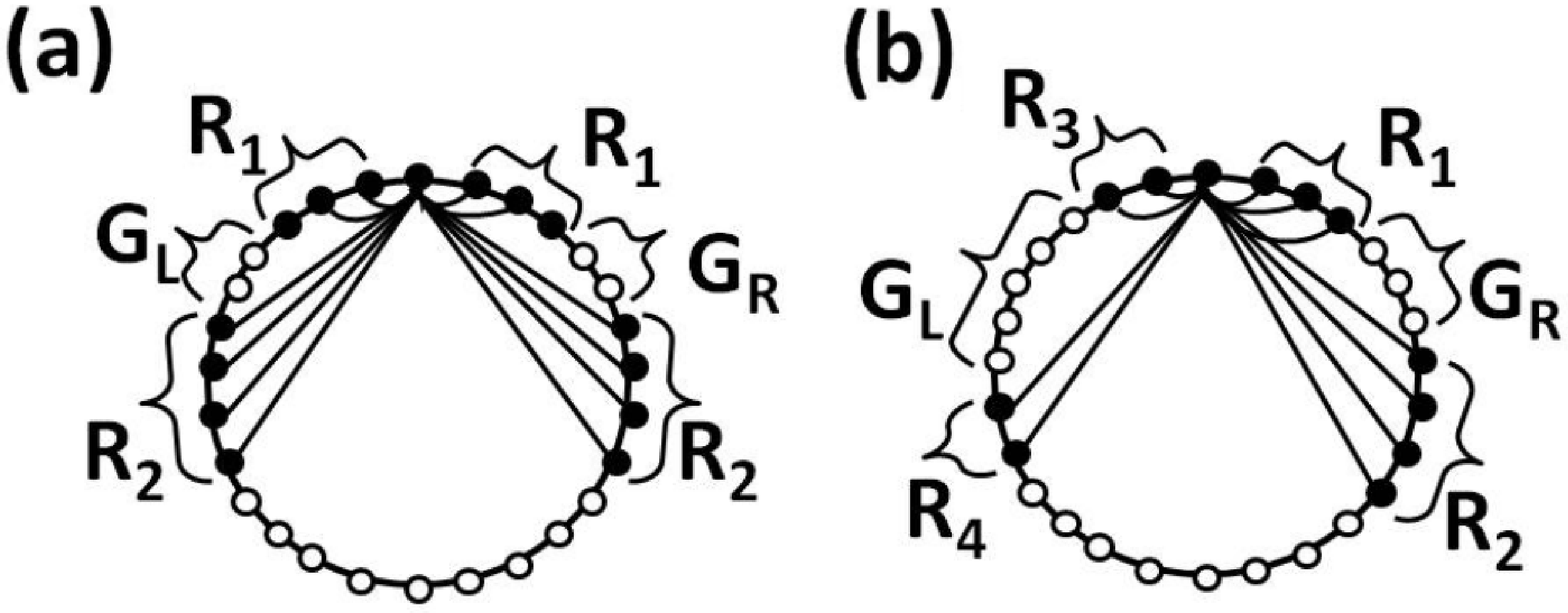}
%\hspace{-0.8cm}\includegraphics[width=0.64\linewidth]{fig6a}\hspace{-1.7cm}\includegraphics[width=0.63\linewidth]{fig6b}
\caption{\label{fig:6}  
Schematic distribution of gaps around a reference node in a ring geometry: (a) symmetrically distributed gaps with $G_L=G_R=G$; (b) asymmetrically distributed gaps. 
%{\bf***ES:  The lettering is too small! Also, all arrows should be reversed since $g_{kl}$ describes input from $l$ to $k$; similarly in Fig.11. Alternatively, we can change $g_{kl}$ to $g_{lk}$ everywhere in the text and in Eq.(5),(6),(9). I suggest the latter, but have not yet done it.***}
}
\end{figure}

Gaps may be introduced in the
connectivity matrix as schematically depicted in Fig.~\ref{fig:6}.
Each oscillator is coupled with a finite number of $2R\le N$
oscillators, $R$ to the left and $R$ to the right. The gaps are arranged
symmetrically or asymmetrically on the left and the right of each node.
If $G_L$ ($G_R$) is the size of a gap to the left (right)
of a node, it follows that $G_L+G_R+2R \le N$. 

In the symmetric case ($G_L=G_R=G$), the adjacency matrix $\{g_{kl}\}$, $k,l=1,\cdots,N$, is given by:
\begin{equation}
 g_{kl} = \left\{
  \begin{array}{l l}
    1 & \quad \text{if $k-R_1<l<k+R_1$}   \\
           & \quad \text{    or $k+R_1+G<l<k+R_1+G+R_2$}\\ 
           & \quad \text{    or $k-R_1-G-R_2<l<k-R_1-G$}\\
    0      & \quad \text{elsewhere},
  \end{array} \right.
\label{eq4-01}
\end{equation}
where all indices are taken modulo $N$.

Each node will have a number of connections $R=R_1+R_2$
symmetrically placed to both its left and right with a number of gaps equal to $G=G_L=G_R$
also symmetrically placed around it [see Fig.~\ref{fig:6}(a)]. Hence, the total number of connections remains $2R$. 
In principle, one can introduce many symmetric gaps on the left and on the right of each element
in a similar way, but this is beyond the scope of this study.

%**********************************************************************************
%**********************************************************************************

In the general case of inhomogeneous gap distributions, the size of the gaps 
as well as the size of the linked regions can be different on the left and the right of each element [see Fig.~\ref{fig:6}(b)], i.e.
\begin{align}
 g_{kl} = \left\{
  \begin{array}{l l}
    1 & \quad \text{if $k-R_3<l<k+R_1$}\\
           & \quad \text{ or  $k+R_1+G_R<l<k+R_1+G_R+R_2$}\\
           & \quad \text{ or  $k-R_3-G_L-R_4<l<k-R_3-G_L$}\\
    0 & \quad \text{elsewhere}.
  \end{array} \right.
\label{eq4-02}
\end{align}
Now, $R_3~(R_1)$ and $R_4~(R_2)$ denote the number of links to the left (right) of the $k-$th element before and after the gap.
Thus, one might have $G_R\ne G_L$.
In the same way, the sizes of the linked regions $R_1$,
$R_2$, $R_3$, and $R_4$ need not be related.
Multiple inhomogeneous gaps may be introduced in a similar way.

Numerical simulations indicate that the presence of gaps tend to stabilize existing chimeras
and to create multi-chimera states. 
As in the previous section, we always start from initial values randomly placed on the
circle of radius $2$, that is $u_k(0)^2+v_k(0)^2=4$.
Each oscillator is coupled with $2R=500$ others, distributed
on equal numbers to the left and to the right, i.e. $R=R_1+R_2=R_3+R_4=250$.
  
%**********************************************************************************  
\begin{figure}[ht!]
\centering
\includegraphics[clip,width=0.60\textwidth,angle=-90]{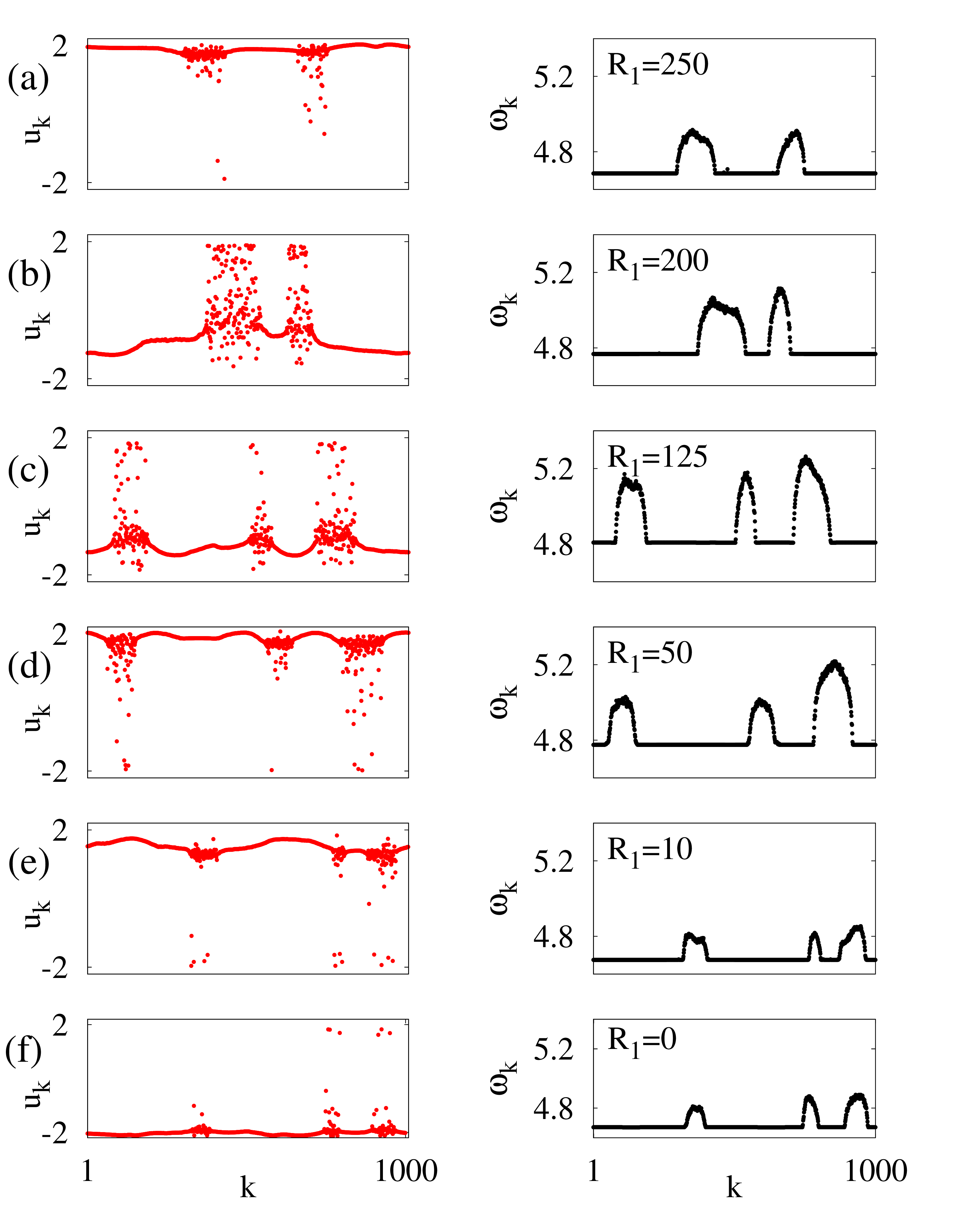}
\caption{\label{fig:7}(Color online) Snapshots of the variables $u_k$ and corresponding mean phase velocities for different gap positions.
Each element is coupled with $2R=500$ others elements
symmetrically distributed around each node.  The connectivity matrix
contains two gaps of size $G_L=G_R=G=100$ symmetrically placed around each node
while the values of $R_1$ and $R_2=R-R_1$ are indicated on each panel. Other parameters: $N=1000, \sigma =0.3, a_k\equiv a=0.5$, and $\eps=0.05$. All simulations start from the same initial conditions.}
\end{figure}
%**********************************************************************************

Figure \ref{fig:7} depicts on the left panels (a)-(f) typical snapshots obtained in
the case of symmetrically distributed gaps, with different values of $R_1$ and $R_2=R-R_1$
 keeping constant the gap sizes $G_L=G_R=G=100$. 
 This way
only the topology of the connectivity changes and not the actual size of the gaps.
The right panels depict the corresponding mean phase velocities and the values
of $R_1$ are indicated.
Starting with the top panels (shapshot (a) and corresponding phase velocities), 
where the case of $R_1=250$, $R_2=0$ is depicted, we recover the case of
$2R=2R_1$ symmetrically distributed links without gaps and a 2-chimera state is
observed, as in Ref.~\cite{OME13}. As the gaps are introduced and moving closer and
closer to the reference node, the mean phase velocities of the (in)coherent regions increase [Fig.~\ref{fig:7}(a) and (b)].
Moving the gaps close to the center, the number of coherent and incoherent
regions changes from 2 to 3. 
It is interesting to note that the difference $\Delta \omega$ in the mean phase velocity between coherent and incoherent
regions becomes maximum when the gaps (on the left and right) split the coupling
regions approximately into equal parts. This will be discussed in greater detail below.

Overall, Fig.~\ref{fig:7} indicates that 
the position of the gaps may change the multiplicity of the chimera
as well as the relative mean phase velocity of coherent and incoherent regions.
This effect depends on the complex interplay between the specific pattern of
the links and the initial conditions.
 
%********************************************************************************** 
\begin{figure}[ht!]

\centering
\includegraphics[clip,width=0.3\textwidth,angle=-90]{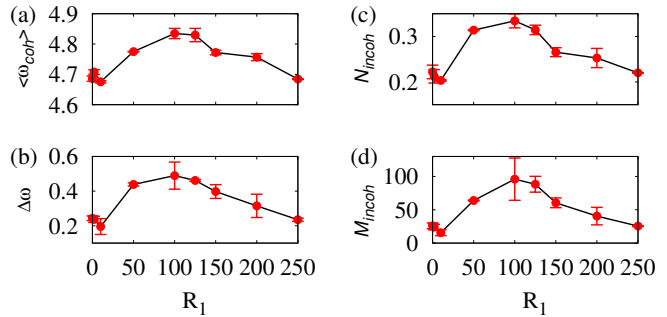}
\caption{\label{fig:8} (Color online) Measures of coherence-incoherence of a chimera state
as a function of the size of the inner coupling regions $R_1$.
Each element is coupled with $2(R_1+R_2)=500$ others elements
symmetrically distributed around each node. 
The connectivity matrix
contains two gaps of size $G_L=G_R=G=100$ symmetrically placed around each node. 
Other parameters as in Fig.~\ref{fig:7}.
Averages are taken over 10 runs.}
\end{figure}
%**********************************************************************************

For a better understanding of the role of the coupling topology for the
development of a chimera state, we
introduce and calculate two measures that account for the relative size of the
coherent versus incoherent part of the chimera.
Let us denote by $\omega_k$ the mean phase velocity of element $k$ and by $\omega_{\text{coh}}$
the mean phase velocity of the coherent parts. Then, the relative size $N_{\text{incoh}}$
of the incoherent
parts of the chimera state is calculated as:
\begin{eqnarray}
 N_{\text{incoh}} = {1\over N}\sum_{k=1}^N\Theta (\omega_k-\omega_{\text{coh}}-c)
\label{eq4-03}
\end{eqnarray}
where $\Theta$ is the step function which takes the value 1 when its argument takes positive
values and zero otherwise. $c$ is a small tolerance, which in this case we set to 0.05.

We also define the extensive cumulative size $M_{\text{incoh}}$ of the incoherent parts as:
 \begin{eqnarray}
 M_{\text{incoh}} = \sum_{k=1}^N |(\omega_k-\omega_{\text{coh}})|.
\label{eq4-04}
\end{eqnarray} 
This is an extensive measure which represents the area below the arcs in the mean phase velocity 
profiles.
$M_{\text{incoh}}$ demonstrates the degree of incoherence of the chimera state and is zero for
purely coherent states.

Figure~\ref{fig:8} displays four measures of coherence 
as a function of the inner coupling regions $R_1$, for the above case of a symmetric
connectivity matrix with $R_1+R_2=250$
elements and two gaps of equal size $G_L=G_R=G=100$ symmetrically placed around the 
reference node: (a) $\langle \omega_{\text{coh}}\rangle$, the mean phase velocity of 
the coherent region, (b) $\Delta \omega$, the maximum difference between incoherent and coherent $\omega$ values,
(c) $N_{\text{incoh}}$, the relative size of incoherent regions, and (d) $M_{\text{incoh}}$,
which provides information about the amount of influence of the incoherent regions. 
We calculate these measures as a function of $R_1$, changing the position of the
gaps symmetrically around the ring. Averages over 10 runs are
shown, starting from different initial conditions randomly chosen on the circle $u_k(0)^2+v_k(0)^2=4$. Averages are necessary to account for stochastic deviations as well as multistability. Error bars in the figure indicate the standard deviations.

Figure~\ref{fig:8}(a) shows that the average mean phase velocity of the coherent parts 
increases as the two gaps move symmetrically away from the reference node
and assumes a maximum around $R_1=125$. At this value the coupling regions before and after
the gap are of equal size.
As the gaps depart further from the
reference node, that is, for further increase of $R_1$, $\langle\omega_{\text{coh}}\rangle$ decreases again to attain the value of the regular ring without gaps for $R_1=250$
and $R_2=0$. 
When $\langle \omega_{\text{coh}}\rangle$ becomes maximum, the difference of the mean phase velocity 
profiles becomes largest [see Fig.~\ref{fig:8}(b)] and so does the relative
number of incoherent oscillators [see Fig.~\ref{fig:8}(c)]. At the same
$R_1$ value, $M_{\text{incoh}}$ also attains its highest value. 

From this discussion it becomes evident that the gaps may
change the position and/or multiplicity of the chimera state. Moreover, the gap position
along the ring
determines the relative ``weight'' of the coherent versus incoherent
part. Therefore, using appropriate gap positions, one may influence the number
of elements belonging to the incoherent parts and also the difference
in the mean phase velocities $\Delta\omega$.

%**********************************************************************************
\begin{figure}[ht!]
\centering
\includegraphics[clip,width=0.32\textwidth,angle=-90]{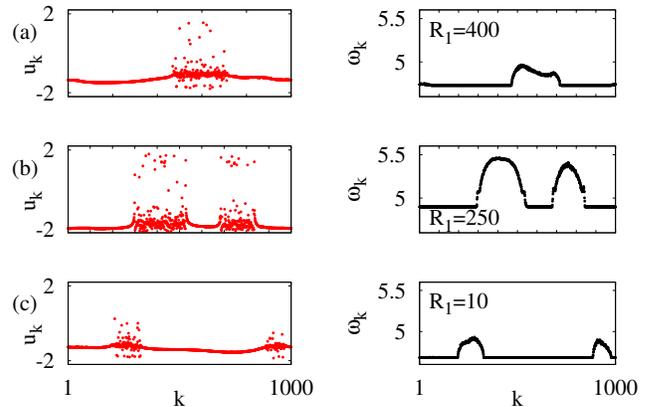}
\caption{\label{fig:9} (Color online) Snapshots and mean phase velocity of the variable $u_k$ for one-sided connectivity.
Each element is coupled with $R=500$ elements to its right, i.e. $R_3=R_4=0$. The connectivity matrix 
is asymmetric and the coupling pattern contains one gap of size $G=100$. The values of $R_1$ and $R_2=R-R_1$ are indicated in the 
panels. Other parameters as in Fig.~\ref{fig:7}.
All realizations start from the same initial conditions.}  
\end{figure}
%**********************************************************************************

Next, we study the question of inhomogeneous gap distribution [Eq.~(\ref{eq4-02})]. In order to avoid introducing too 
many parameters in the system, we keep $R_3=R_4=G_L=0$, thus
allowing links only on the right of each element. We consider one gap of size $G_R=100$
and various values of $R_1$ and $R_2$ under the condition $R_1+R_2=R=500$. This keeps the same number of links as above, for
comparison with the previous case of the symmetrically placed gaps.
Due to the asymmetry in the coupling, the network becomes directed: if a node
$k$ receives a signal from a node $l$, the opposite does not necessarily hold as in the
symmetric case.
In addition, the reference node is not always centrally located as in the symmetric case.

Figure~\ref{fig:9} depicts typical $u_k$ snapshots and corresponding mean phase velocity profiles of asymmetric connectivity matrices with one gap for different
values of the parameters $R_1$ and $R_2$, while the sum $R_1+R_2=R$ is kept constant, $R=500$.
Phenomena of splitting and merging of the incoherent parts are typically 
observed for different spatial arrangements of the gap. 
By shifting the gap towards the
reference node, the difference
in the mean phase velocity profile attains maximum values when the
gap separates the coupling regions into equal parts. As the gap moves further away from the
reference node, the mean phase velocity decreases again. This behavior is verified for
different values of $R$ and for $G$.

%**********************************************************************************
\begin{figure}[ht!]
\centering
\includegraphics[clip,width=0.27\textwidth,angle=-90]{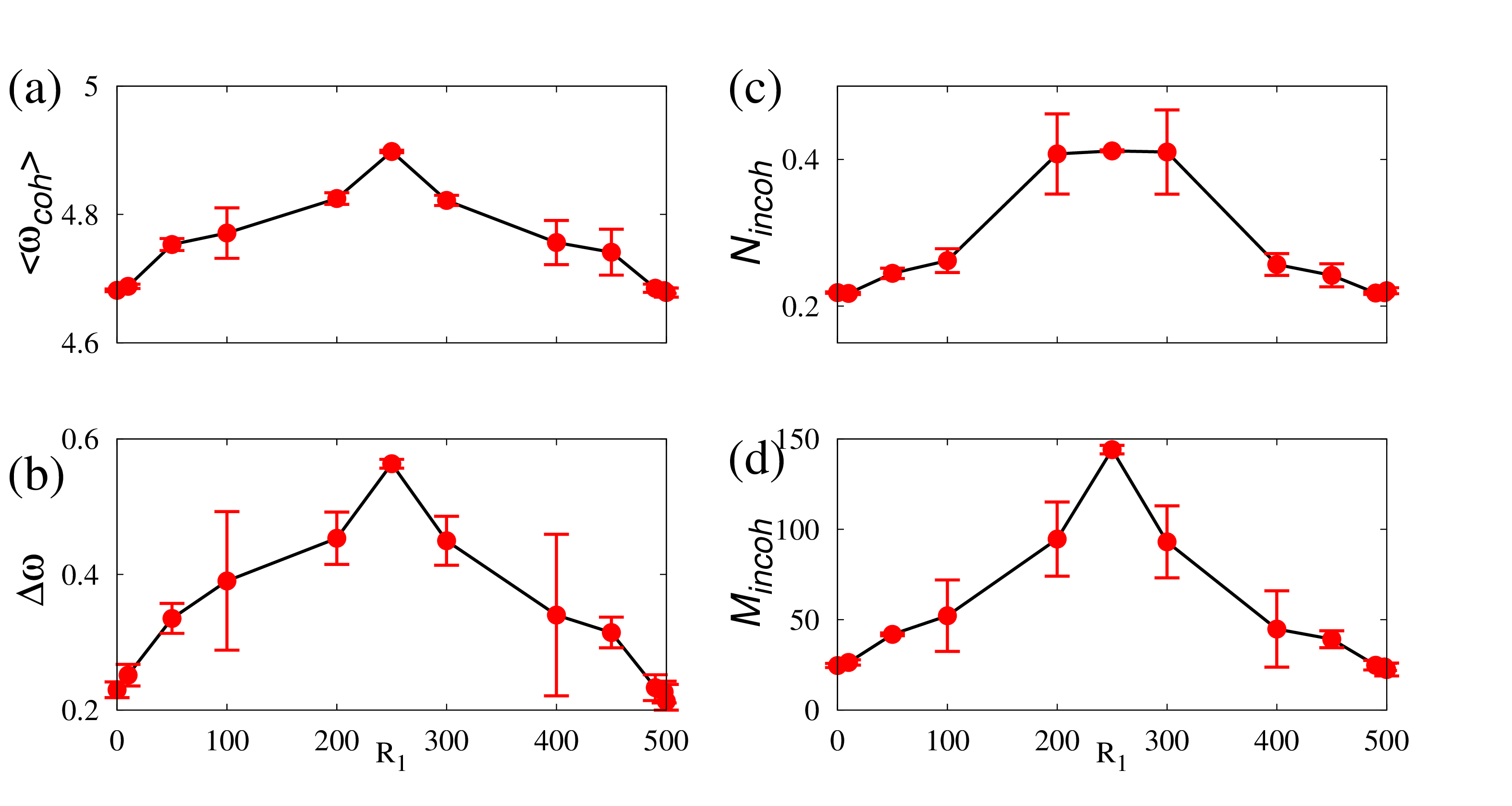}
\caption{\label{fig:10} (Color online)  Measures of coherence-incoherence of a chimera state.
Each element is coupled with $R=R_1+R_2=500$ elements to its right. The connectivity matrix 
is asymmetric and
contains one gap of size $G=100$. All other parameters as in Fig. \ref{fig:9}.
Averages are taken over 10 runs.} 
\end{figure}
%**********************************************************************************

To investigate 
further the properties of coherent-incoherent
regions as a function of the position of the gap with respect to the reference node, we
calculate the four measures that account for the presence of chimera states, in analogy to the
case of symmetrically placed gaps shown in Fig.~\ref{fig:8}. In the simulations, all other parameters are kept constant,
as in Fig.~\ref{fig:9}, while varying the size of the inner and outer coupling radii $R_1$ and 
$R_2$, retaining their sum constant, $R_1+R_2=500$. For each parameter pair $(R_1,R_2)$ 
averages were taken over 10 realizations starting from different initial conditions, to take into account stochastic effects and
multistability. The calculations of the various measures of coherence are shown in Fig.~\ref{fig:10}
as a function of $R_1$. 

It is interesting to note that as in the case of symmetric connectivity (Fig.~\ref{fig:8}), both the mean phase velocity 
$\langle \omega_{\text{coh}}\rangle$ of the coherent regions and the maximum $\Delta \omega$ is achieved when the gaps are placed centrally, separating the coupling regions in approximately
equal parts [Fig.~\ref{fig:10}(a) and (b)]. The same observation holds for the other two measures, $N_{\text{incoh}}$ and 
$M_{\text{incoh}}$.
Therefore, the introduction of gaps can be used as a control parameter to modulate 
the synchronization and the magnitude of phase velocities in the coherent and incoherent
parts of the chimera states.  
The above results are indicative of the complexity
induced in the chimera states by the spatial arrangement of the connections
even if all the other system parameters are kept fixed.

\section{Hierarchical Gap Distribution and Chimeras}
\label{sec:fractal}

In the previous section, we have presented first evidence
 that the presence and the position of gaps 
may modify the coherence properties of the chimera state and its multiplicity. 
In the simulations, the position and the size 
of the gaps did not follow any precise distribution, but was specifically designed
to exemplarily demonstrate the changes in the chimera properties with the gap position and symmetry.
We now proceed to consider distributions of the gaps with specific hierarchical architecture.
The study of different hierarchical architectures in the neuron connectivity is motivated
by MRI results of the brain structure which show that the neuron axons
network spans the brain area fractally and not homogeneously  
\cite{KAT09,EXP11,KAT12}. In the remain of this
section the word 'fractal' will be employed to denote
mainly hierarchical structures of finite orders $n$, since the
human brain has finite size and does not cover all orders,
$n \to \infty$ (as in the exact definition of a fractal set).

The fractal connectivity dictates a hierarchical ordering in the distribution of neurons
which is essential for the fast and optimal handling of information in the brain.
When we study phenomena, which involve coupled neurons, the need to take into account the fractal architecture of the
neuron network becomes apparent. This is a problem that has emerged after the recent development of 
DT-MRI techniques allowing for the detection of the neuron network
architecture with high levels of accuracy, while traditionally the synchronization
properties of coupled neurons are studied in ring architectures with local, nonlocal, or global couplings. 

Simple hierarchical structures can be constructed using the 
classic Cantor fractal construction
process \cite{MAN83,FED88}. Using the iterative bottom-up procedure
to construct the Cantor set, we create a symbolic sequence consisting of two symbols hierarchically nested in one another. Starting with a base containing $b$ symbols
(0's or 1's) we iterate it a number of times $n$ and thus obtain systems of size
 $N=b^n$. By closing this string of $b^n$ symbols in a ring we construct 
a hierarchical connectivity matrix $g^{\{ n\} }_{kl}$ considering that the symbol 1 denotes the existence
of a link, while the symbol ``0'' the absence of a link, namely,
\begin{align}
 g^{\{ n\} }_{kl} = \left\{
  \begin{array}{l l}
    1 & \quad \text{if both nodes $k$ and $l$ belong to the}\\
           & \quad \text{ Cantor set ($n-$th iteration)}\\ \\
    0 & \quad \text{elsewhere}.
  \end{array} \right.
\label{eq4-05}
\end{align}

\begin{figure}[ht!]
%\begin{centering}
\centering
\includegraphics[width=0.75\linewidth]{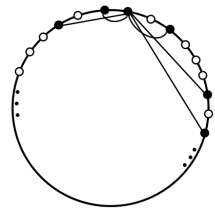}
%\includegraphics[clip,width=\linewidth]{fig11}
% \includegraphics[clip,width=0.3\textwidth,angle=0]{fig11.eps}
%\end{centering}
\caption{\label{fig:11} Schematic distributions of hierarchical gaps in a ring geometry
based on the triadic (101) Cantor set.}
\end{figure}

In this way a connectivity 
matrix  of size $N=b^n$ is constructed, which contains a hierarchical distribution of
gaps with a variety of sizes. The number of times the symbol ``1''
appears  in the base, denoted by $c_1$, defines formally the fractal dimension 
$d_f$ of the structure, as $d_f=\ln c_1 /\ln b$. This measure $d_f$
describes perfectly the fractal structure when the number of iterations $n\to \infty$,
while for $n\ll\infty$ the structure is called hierarchical, because it has been
constructed based on a hierarchical algorithm. In Fig.~\ref{fig:11} we present
schematically the construction of a ring connectivity using the triadic (101)
Cantor set~\cite{FED88}.

%**********************************************************************************
%**********************************************************************************

In the current study we use a construction
base $b=6$. Thus, the size of the network is given in powers of $6$. For comparison with previous results we use the $n=4$ iteration of the fractal 
giving a total size of the network $N=b^4=1296$ nodes. We
change the value of $c_1$ (number of times the symbol ``1'' 
is encountered within the base $b$) to vary the number of hierarchically coupled elements
in the structure and thus to vary the fractal dimension of the structure. Fixing $c_1$ still leaves one degree of freedom, that is, the position of the symbols ``1'' in the base pattern. This issue will be addressed later.
All other parameters, 
apart from the number and position of the coupled elements, are kept
constant as in the previous section and Figs.~\ref{fig:7}-\ref{fig:10}.

%**********************************************************************************
\begin{figure}[ht!]
%\begin{centering}
\includegraphics[clip,width=0.35\textwidth,angle=-90]{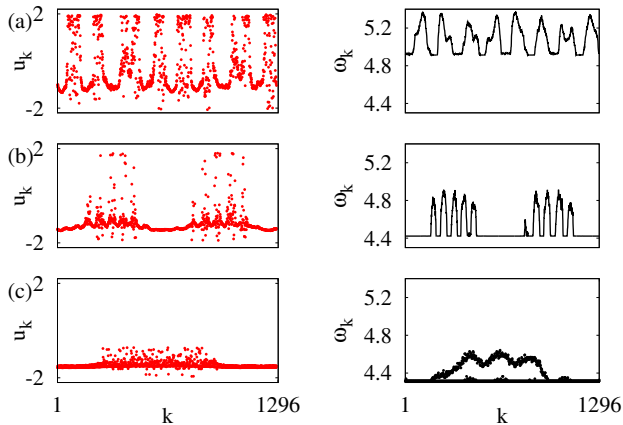}
%\end{centering}
\caption{\label{fig:12} (Color online) Snapshots of the variable $u_k$  and corresponding mean phase
velocities $\omega _k$ for different hierarchical
connectivity matrices.
 The fractal dimensions of the hierarchical structures are: (a)~$d_f=\ln 3/\ln 6=0.6132$ (base pattern: $100101$),
(b)~$d_f=\ln 4/\ln 6=0.7737$ (base pattern: $101110$) and (c)~$d_f=\ln 5/\ln 6=0.8982$ (base pattern: $110111$). Parameters: $N=6^4=1296$ and other parameters as in Fig.~\ref{fig:7}. 
}

\end{figure}
%**********************************************************************************

In Fig.~\ref{fig:12}(a) the calculations are performed with $c_1=3$. The connectivity matrix 
is very sparse and has a fractal dimension $d_f=\ln 3/\ln 6=0.6132$.  In the $n=4$ iteration it
contains only $c_1^n=3^4=81$ links of each element with the others, each element failing to link
with the remaining $b^n-c_1^n=1215$ elements. The links are mostly isolated
while the gaps cover most of the structure. A multi-chimera with multiplicity $8$ and clearly
identified coherent/incoherent parts is observed. This result agrees with previously published
works indicating that the multiplicity of a chimera state is high when the number of
links in the ring network is small~\cite{VUE14a}. 

In Fig.~\ref{fig:12}(b) the number of links was increased, using $c_1=4$. 
The connectivity matrix 
has a fractal dimension $d_f=\ln 4/\ln 6=0.7737$ and each element is connected to $c_1^n=256$
others. Here, the chimera represents a nested structure, containing 10 coherent/incoherent regions
clustered into two parts. The two incoherent parts show a substructure consisting of five closely packed incoherent regions.  
The two clusters are separated by large coherent regions. This phenomenon will
be further explored below. 
To continue with the dependence of the qualitative and
quantitative features of the chimera state on the hierarchical architecture, in Fig.~\ref{fig:12}(c)
we consider 
$c_1=5$. The number of links increases further and now each element is coupled with $c_1^n=5^4=625$
elements. The incoherent parts seem to merge into a $1$-chimera, 
but the calculation of the phase velocity demonstrates that this single incoherent region has
a substructure with three maxima.

The current results support and extend those of Ref.~\cite{OME13}, where it was demonstrated
that the increase in the number of links leads to decrease in the chimera multiplicity.
In addition, here we give first evidence that if the links are distributed in a
hierarchical manner, the corresponding chimera state shows nested complex 
incoherent patterns. Similar structure of chimera states were found in systems of phase oscillators with repulsive coupling~\cite{MAI14}.

%**********************************************************************************
\begin{figure*}[ht!]
\includegraphics[width=0.3\linewidth,angle=-90]{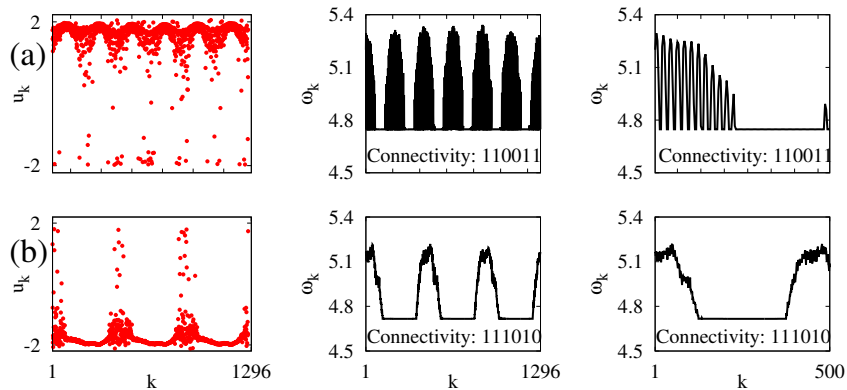}
\caption{\label{fig:13} (Color online) Snapshots of the the variable $u_k$ (on the left)
 corresponding
mean phase velocities $\omega_k$ (middle) and details of mean phase velocities (on the right) for different Cantor patterns.
The fractal dimensions in all hierarchical structures is constant $d_f=\ln 4/\ln 6$. 
The initiation strings are : (a) 110011 and  (b) 111010. The left panels are zooms of the corresponding mean phase velocities 
(middle panels) 
 to demonstrate structural ramifications. 
Parameters as in Fig.~\ref{fig:12}.}
\end{figure*}
%**********************************************************************************

Next, we investigate the influence of the local structure of the connectivity matrix
without changing the fractal dimension. In other words, we only 
change the local configuration of the fractal. As an example, we use two different fractal
realizations that have the same dimension $d_f=\ln 4/\ln 6$, but start from different initialization strings (base patterns): 
(a)~the string 110011 (symmetric) and (b)~the string 111010 (asymmetric). The results are presented in
Fig.~\ref{fig:13}.

In particular, Fig.~\ref{fig:13}(a) depicts the chimera state when the connectivity matrix is
initiated by the string 110011. The corresponding mean phase velocity profile (middle panel)
shows six incoherent
regions. The right panel of Fig.~\ref{fig:13}(a) depicts an enlargement for the first 150 oscillators.
From this panel, it is evident that the structure is ramified and that within
each incoherent regions there are nested coherent and incoherent ones. This is certainly
reminiscent of the hierarchical structure of the connectivity matrix which includes
nested gaps between linked elements. 

We now turn to the second example, the initiation string
111010. The fourth iteration of this string results in producing  a connectivity matrix with
multiple sizes of gaps and solid interacting regions around each reference element.
The resulting picture (snapshot) of the amplitude of the oscillators, Figs.~\ref{fig:13}(b),
demonstrates the presence of a chimera state with fewer coherent/incoherent
regions. Zooming into the incoherent regions [Fig.~\ref{fig:13}(b), right panel]
does not demonstrate any trace of ramifications. This picture suggests that
(i) the fractal dimension is not the only parameter which determines the multiplicity
or the pattern of the chimera state, and (ii) different initiation strings, producing
the same fractal dimension,  give rise to different chimera patterns.

These observations allow us to propose that the presence or absence of ramifications in Fig.~\ref{fig:13}
may be attributed to the presence of symmetries in the pattern of the original string.
If the size of the system is increased by using higher iteration levels in the
connectivity matrix, then the hierarchical structure will be more evident and
then nested structures may arise even for original strings which do not
present obvious symmetries. This hypothesis needs to be tested with extended
simulations in future publications.

\begin{figure}
\includegraphics[clip,width=0.57\textwidth,angle=-90]{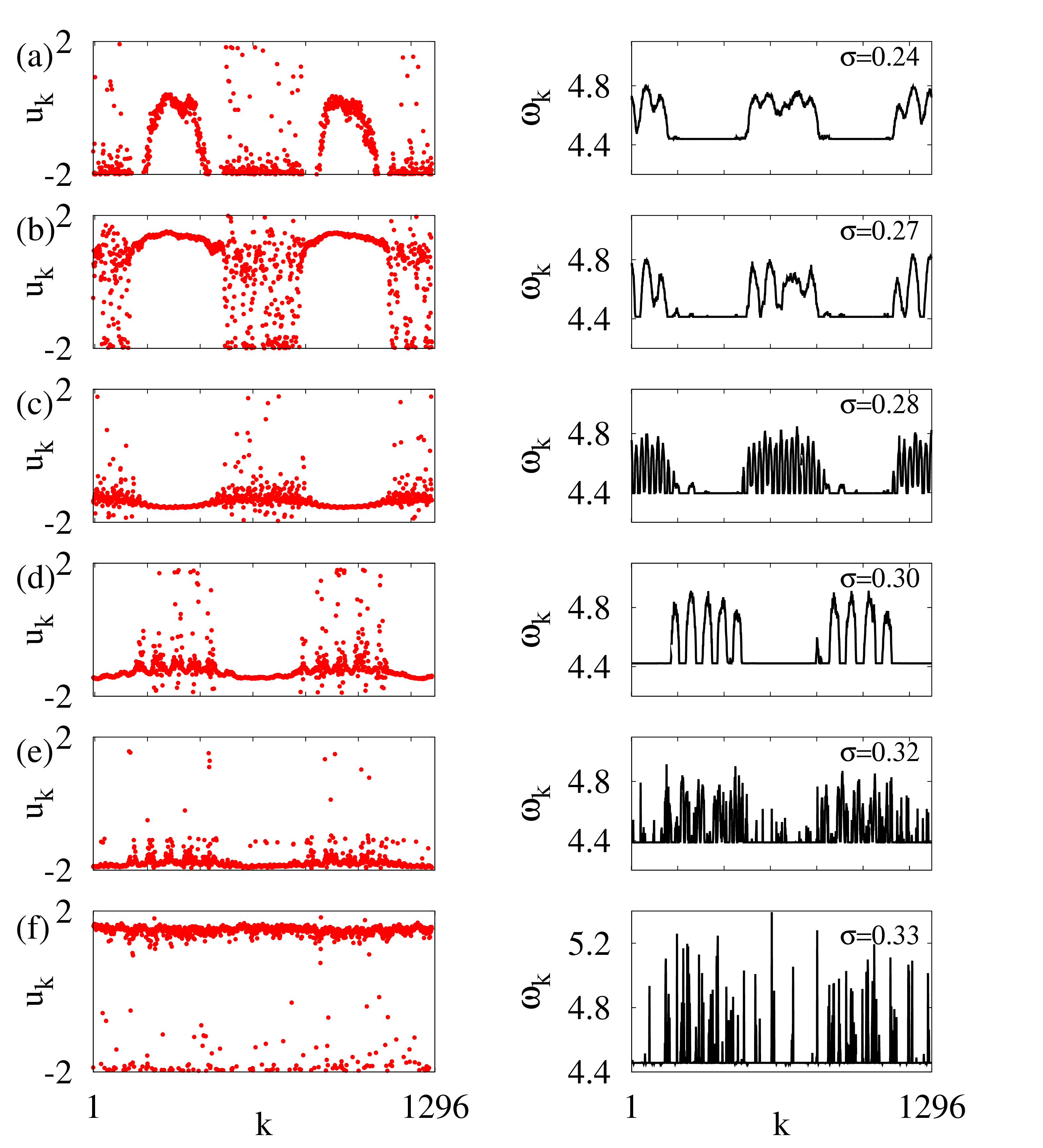}
\caption{\label{fig:14} (Color online) Snapshots of the the variable $u_k$ and the
corresponding mean phase velocities $\omega_k$ for different coupling strengths.
 The fractal dimensions of the hierarchical structure is constant $d_f=\ln 4/\ln 6$. The
panels correspond to: (a) 
 $\sigma =0.24$,
(b) $\sigma =0.27$, (c) $\sigma =0.28$
(d) $\sigma =0.30$ [same as Fig. \ref{fig:12}(b)],
(e) $\sigma =0.32$, (f) $\sigma =0.33$. Initiation string is $101110$ and other parameters are
as in Fig.~\ref{fig:12}.}
\end{figure}

Finally, we investigate the change of coherence in the chimera state 
when we change the coupling strength $\sigma$.
We keep the fractal dimension fixed at $d_f=\ln 4/\ln 6$ and consider four iterations. Thus, the coupling architecture contains $4^4=256$ hierarchically nested links. Figure~\ref{fig:14} shows how the
chimera multiplicity changes with $\sigma$. In fact, multiplicity does not change, but it becomes ramified.
Starting from $\sigma =0.24$, two well-defined coherent (and two incoherent) regions develop. The same
picture continues for $\sigma = 0.27$, but the coherent regions slowly acquire a finer structure
inside, which can be better quantified by the mean phase velocity profiles $\omega_k$ 
(see Fig.~\ref{fig:14}, right panels). 
The fine structures
become best developed for $\sigma =0.30$, while they slowly decay
 for larger $\sigma$ values. Note that even for $\sigma =0.24-0.27$ we may observe
a wavy structure in the mean phase velocities around the maxima of the incoherent regions,
which is not observed in the well formed arc shape of the classical chimeras (see
Figs.~\ref{fig2} and \ref{fig3}). Therefore, the coupling strength $\sigma$ acts as a control parameter
for turning the ramifications on and off.

These findings constitute some first observations of ramified structures emerging when
the link architectures are not solid but contain nested, hierarchical gaps. It is 
important to add here that for a given system size it is not possible to predict the 
size and number of coherent and incoherent gaps, solely from the fractal dimension 
and/or the number of links in the system. The interplay between the hierarchical geometry 
of the links and the randomness of the initial conditions may lead to different chimera
profiles as was seen in Fig.~\ref{fig:13} where in cases (a) and (b) all parameter
were identical including identical initial conditions and number of links, the only 
difference being in the spatial/hierarchical arrangement of the links. For this reason
it is not helpful to study the chimera statistics, but rather one needs to address
the basin of attraction that leads to specific chimera patterns.

The general conclusion of this section is that connectivity matrices, which
contain hierarchically nested gaps around the reference element, 
commonly induce nested coherent and incoherent
regions. These nested structures are best demonstrated by the
mean phase velocity diagrams.
In particular, for the same fractal dimension the structure of the ramifications in the incoherent
regions is tuned by the value of $\sigma$, and by the pattern of
 the initial string used to construct the connectivity matrix in connection 
with the initial conditions.

%**********************************************************************************
%**********************************************************************************

\section{Conclusions}
\label{sec:concl}
In the current study, we have addressed the issue of robustness of chimera states in systems of nonlocally coupled units 
with respect to perturbations of the frequencies of individual elements, and structural transformations of the network topology. We demonstrated that in the system of nonidentical FitzHugh-Nagumo oscillators with regular nonlocal coupling topology, chimera states are robust for small inhomogeneity. As the inhomogeneity increases, chimera  states with multiple incoherent regions transform into chimera states with one incoherent region. 
This finding could be important from the point of view of experiments and applications. In real-world situations, one hardly finds a system with absolutely identical elements.  

Random structural perturbations of the network topology do not destroy the chimera states immediately, and for relatively small numbers of random links, chimera states, as well as their stability regions, can be determined. We have compared the two cases when already existing links in the nonlocally coupled network are either rewired or kept, and provide a statistical analysis of chimera occurrences as the irregularity of the network topology increases. Chimera states might still be realized 
even for a large number of random links.

The introduction of connectivity gaps has allowed us to analyze the symmetric and asymmetric network topologies, when gaps are on  both or only on one side of the reference node, respectively. We have demonstrated numerically that the position of the gaps  
influences the multiplicity and the relative weight of the incoherent and coherent regions of chimera states. Moreover, when the gaps separate the connected areas into equal parts, the difference between the mean phase velocities of the coherent
and incoherent domains attains a maximum. 

In the case of multiple connectivity gaps, we have shown that when the gap distribution is hierarchical, the resulting chimera
pattern exhibits nested incoherent parts. The specific structure depends on the complex interplay between the hierarchical link geometry and the initial conditions.

Our findings corroborate the universal existence of chimera states and extend the variety of systems where chimera states can be found. These results can be useful from the point of view of applications, where inhomogeneity of the elements, and more complex coupling topologies are common.

\begin{acknowledgments}
 This work was supported by the German Academic Exchange Service DAAD and the Greek State Scholarship Foundation IKY within the PPP-IKYDA framework. 
 IO and PH acknowledge support by BMBF (grant no.  01Q1001B) in the framework of BCCN Berlin (Project A13). 
 ES, IO, and PH acknowledge support by Deutsche Forschungsgemeinschaft in the framework of SFB 910. 
The project was also funded by the John S. Latsis Public Benefit Foundation.
\end{acknowledgments}

%\bibliography{ref}
%\bibliographystyle{prsty-fullauthor}

\end{document}